%% file: ttv4.tex
\documentclass{emulateapj}
\usepackage{color}
\usepackage{latexsym, graphicx, amssymb, longtable, epsf}
\usepackage{CJK}
\newcommand{\rr}[1]{{\textcolor{black}{#1}}}

\newcommand{\rev}[1]{{#1}}

\long\def\symbolfootnote[#1]#2{\begingroup\def\thefootnote{\fnsymbol{footnote}}
\footnote[#1]{#2}\endgroup}

\shorttitle{Thirty Planets}
\shortauthors{Xie}

\begin{document}

\begin{CJK*}{UTF8}{gbsn}

\title{Transit Timing Variation of Near-Resonance Planetary Pairs. II.  Confirmation of 30 planets in 15 Multiple Planet Systems}
\author{Ji-Wei Xie (谢基伟)$^{1, 2}$}

\affil{$^1$Department of Astronomy \& Key Laboratory of Modern Astronomy and Astrophysics in Ministry of Education, Nanjing University, Nanjing, 210093, China; jwxie@nju.edu.cn}
\affil{$^2$Department of Astronomy and Astrophysics, University of Toronto, Toronto, ON M5S 3H4, Canada; jwxie@astro.utoronto.ca}


\begin{abstract}
Following on from Paper I in our series \citep{Xie13}, we report the confirmation by Transit Timing Variations (TTVs) of a further 30 planets in 15 multiple planet systems, using the publicly available
 \emph{Kepler} light curves (Q0-Q16).
All of these fifteen pairs are near first-order Mean Motion Resonances (MMR), showing sinusoidal TTVs consistent with theoretically predicted periods, which demonstrate they are orbiting and interacting in the same systems.  Although individual masses cannot be accurately extracted based only on TTVs (because of the well known degeneracy between mass and eccentricity), the measured TTV phases and amplitudes can still place \emph{relatively} tight constraints on their mass ratios and upper limits on their masses, which confirm their planetary nature.   Some of these systems (KOI-274, KOI-285, KOI-370 and KOI-2672) are relatively bright  and thus suitable for further follow-up observations.
\end{abstract}



\section{Introduction}
As of today, the \emph{Kepler} mission has found a few thousand  planetary candidates \citep{Bat12, OD12, Hua13, Bur13}, and about 140 of them have been confirmed as bona fide planets\footnote{http://exoplanetarchive.ipac.caltech.edu}. 
Traditional methods of planet confirmation, relying on radial velocity (RV) measurements, detailed dynamical modelling of Transit Timing Variations \citep[TTV, e.g.,][]{HM05, Ago05, Hol10, Lis11}, and/or other complementary data \citep[e.g. BLENDER;][]{Fre11, Tor11}, are time consuming for individual systems and thus not practical for the thousands of Kepler candidates. More recently, some efficient TTV methods for the confirmation of planets have been developed and proven to be productive \citep{For12a, Ste12, Fab12a,  Xie13, Ste13}.

One of these efficient methods uses anti-correlated TTVs to demonstrate that the planetary candidates are orbiting the same star \citep{For12a, Ste12, Ste13}, and then places upper limits on their masses to confirm their planetary nature based on the constraints from dynamical stability \citep{For12a, Ste12, Fab12a,  Ste13}.  

In the first paper of our series \citep[][hereafter Paper I]{Xie13},  we developed a complementary method for the efficient confirmation of planets using TTVs. Specifically, we first demonstrate that a pair of planets are orbiting in the same system  by their sinusoidal TTVs with theoretically predicted periods (this part is similar to that of \citet{Fab12a}). Then based on the theoretical TTV formulas \citep{Lit12}, hereafter LXW12, we use their TTV phases and amplitudes to place upper limits on their masses, confirming their planetary nature. Using this method, we confirmed 24 planets in 12 systems in Paper I.

Here, in the second paper in our series, we continue with this methodology, confirming 30 planets in 15 systems using Kepler's four year light curves (Q0-Q16)\footnote{We note that during the preparation of this manuscript, several of the systems that report here (KOI-1236, KOI-1563 and KOI-2672) were independently confirmed using  a different methodology in the recent publication by \citet{Min13}.}. This paper is organized as follows. In section 2, we present the confirmation of these 30 planets. Section 3 is the summary.

\section{Confirmation of 30 Planets with TTV}

\subsection{Transit time measurements}

The data used in this paper are the long cadence (LC), ``corrected'' light curves (PDC) of KOIs from Q0 to Q16, which are available at Multi-mission Archive at STScI (MAST{\footnote{http://archive.stsci.edu/kepler}}). 
The pipeline used to measure the transit times has been developed and
described in Paper I. In addition, we discard a handful of transit time measurements due to their very low individual transit signal noise ratios, SNR $<$ 2.5. This causes KOI-274.02  and KOI-1873.01 to have TTV data length significantly shorter than their counterparts (Fig.\ref{fig:ttv1} and \ref{fig:ttv3}).   
Our transit time measurements are listed in table \ref{tab:ttvdata} and also publicly available at http://www.astro.utoronto.ca/$\sim$jwxie/TTV. We note that some systems studied here are also listed in the TTV catalog of \citet{Maz13}, which measures the transit times using Kepler data from Q0 to Q12. We compared our transit time measurements to theirs and found good consistency. 

\begin{figure}
\begin{center}
\includegraphics[width=\textwidth]{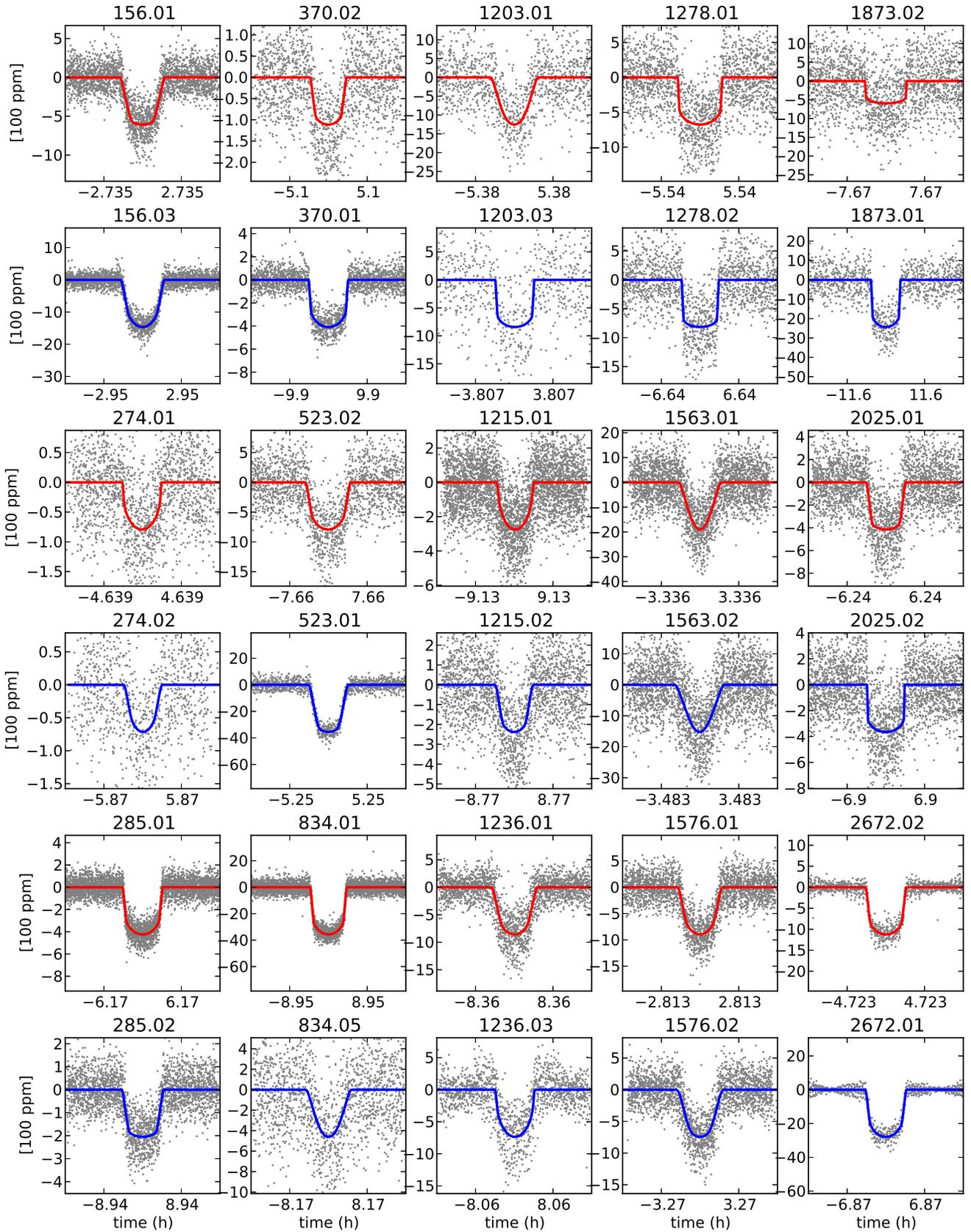}
\caption{Folded light curves (superposition of all transit light-curve segments by setting all central transit times = 0) for the 15 pairs of newly confirmed planets. On the top of the light curves (solid points), there are red and blue solid lines showing the best transit model fits to the inner and outer planet of each pair, respectively. Note the different scales on the horizontal and vertical axes}
\label{fig:template}
   \end{center}
\end{figure}

One by-product of measuring transit times is the folded light curve, from which one can see the significance of the transit. In Figure \ref{fig:template}, we plot the folded light curves for these 30 newly confirmed planets as well as the best MCMC \citep{Pat10} fit using the analytical transit model given by \citet{MA02}.

 \subsection{TTV phase and amplitude}
As in Paper I, we focus on planet candidate pairs which are near first order MMR, 
i.e.
\begin{eqnarray}
\frac{P^{'}}  {P}  \sim \frac{j}{j-1}
\end{eqnarray}
where $j=2,3, . . .$,  $P$ and $P^{'}$ are the orbital periods of the inner and outer candidates, respectively.
Throughout this paper, we adopt the following convention: properties of the outer (inner) candidate in a pair are denoted with (without) a superscript  ``$'$ ''.
The proximity to resonance is defined as
\begin{eqnarray}
\Delta = \frac{P^{'}}  {P} \frac{j-1}{j} -1 .
\end{eqnarray}

  As derived in LXW12, for two transiting objects near first order MMR, the series of transit times both consist of curves with combined linear and sinusoidal components, whose forms can be expressed as: 
 \begin{eqnarray}
 t = T + P \times n + |V|{\rm sin}(\lambda^{j} + \angle V), \nonumber \\
 t^{'} = T^{'} + P^{'} \times n^{'} + |V^{'}|{\rm sin}(\lambda^{j} + \angle V^{'}),
 \label{ttveqn}
\end{eqnarray}
where $t$ and $t^{'}$,  $n$ and $n^{'}$, $T$ and  $T^{'}$, $P$ and $P^{'}$ and  $V$ and $V^{'}$ are the transit time,  transit sequence number, transit epoch, transit period, and TTV complex (see Eqn.7, $\angle V$ denotes the phase of complex $V$) of the inner and outer planets, respectively, and the \emph{longitude of conjunctions}, 
\begin{eqnarray}
\lambda^{j} & = & j\lambda^{'}-(j-1)\lambda \nonumber \\
                     & = & -\frac{2\pi}{P^{'}/j\Delta} (t-T^{'}) + \frac{2\pi}{P} (j-1)(T-T^{'}),
\end{eqnarray}
where $\lambda = \frac{2\pi}{P}(t-T)$ and $\lambda^{'} = \frac{2\pi}{P^{'}}(t-T^{'})$ are the mean longitudes of the inner and outer planet, respectively. The TTV period, which we refer to as the \emph{super-period}, is 
\begin{eqnarray}
P^{j}=\frac{P^{'}}{j|\Delta|},
\label{Pj}
\end{eqnarray}
and 
\begin{eqnarray}
A_{\rm ttv}  =   |V|, \,\,  & A^{'}_{\rm ttv}  =   |V^{'}| \nonumber \\
\phi_{\rm ttv} =  \angle \left(\frac{\Delta}{|\Delta|}V\right), \,\, &  \phi^{'}_{\rm ttv} = \angle \left(\frac{\Delta}{|\Delta|}V^{'}\right)
\label{amp_pha}
\end{eqnarray}
are the TTV amplitudes and phases, respectively.

We use a MCMC method \citep{Pat10} to fit the above model (Eqn.\ref{ttveqn}) to the measured transit times.
Figures \ref{fig:ttv1}, \ref{fig:ttv2} and \ref{fig:ttv3} plot the TTV data (residuals from the linear fit) and best fits for the 15 newly confirmed planet pairs. The fitting results are summarized in table \ref{tab:ttvfit}.

 \subsection{FAP Analysis} 
Figures \ref{fig:ttv1}-\ref{fig:ttv3} illustrate the 15 KOI pairs in our investigation, comparing their measured TTVs with the theoretically predicted period (super-period).   Nevertheless, one should consider how likely that such a pair of TTVs could be produced as an artefact of the noise in the data rather that from a real signal arising from a pair of interacting planets, i.e. false alarm probability (FAP). 
To address this issue, Paper I proposed two methods; one is based on the Lomb-Scargle (LS) periodogram \citep{Sca82, ZK09} and the other is based on data refitting (similar to that in \citep{Fab12a}). 
In fact, these two methods are intrinsically equivalent although the first one is faster. Thus, here we only adopt the first method to calculate the FAP. 
Specifically,  for each pair of measured TTVs, we first compute the LS periodogram and record their powers at the super-period. We then calculate the LS powers at the super-period for another $10^{4}$ sets of random permutations of the original TTV data.  The FAP is estimated as the fraction of realization pairs with both LS powers larger than that of the corresponding original TTVs \footnote{\rr{We note that such a method underestimates the FAP by a factor of two if both TTV data are pure noise. However, none of the systems confirmed in this paper applies to this extreme case.}}.   
 
 The results of the above FAP analysis are listed in table \ref{tab:ttvfit}. All of the TTV pairs have a FAP less than 0.3\%, corresponding to a confidence level over 99.7\% (3-$\sigma$). Such low FAPs mean that we can be highly confident that each pair of objects are orbiting the same star. 
 
\subsection{Mass constraints from TTV}
\label{sec:mass}
 As derived in LXW12, the TTV amplitudes and phases (or complex TTV) explicitly reveal the masses and eccentricity of the system, 
\begin{eqnarray}
V  & = &  P \frac{\mu^{'}}{\pi j^{2/3}(j-1)^{1/3}\Delta}\left(-f-\frac{3}{2}\frac{Z^{*}_{\rm free}}{\Delta}\right) \nonumber \\
V^{'}  & = &  P^{'} \frac{\mu}{\pi j\Delta}\left(-g+\frac{3}{2}\frac{Z^{*}_{\rm free}}{\Delta} \right),
\label{vv}
\end{eqnarray}
where $\mu$ and $\mu^{'}$ are the mass ratio of the inner and outer objects to the star, respectively, 
$f$ and $g$ are sums of Laplace coefficients (of order of unity), as listed in Table A1 of LXW12, and $Z^{*}_{\rm free}$ is the complex conjugate of 
$Z_{\rm free}=f z_{\rm free} + g z^{'}_{\rm free}$, a linear combination of the free complex eccentricities of the two planets.  

In principle,  $\mu$, $\mu^{'}$ and $Z_{\rm free}$ (both real and imaginary parts) can be inferred by inverting Eqn.\ref{vv}. In reality, however, $Z_{\rm free}$ is highly degenerate with $\mu$ and/or $\mu^{'}$ (see below and also in LXW12 and Paper I). Nevertheless, as discussed in LXW12,  one can derive the nominal masses from Eqn.\ref{vv} by assuming $|Z_{\rm free}|=0$, i.e.,
\begin{eqnarray}
m_{\rm nom} &=& M_{\star}\left|\frac{V^{'}\Delta}{P^{'}g}\right|\pi j \nonumber \\
m^{'}_{\rm nom}& =& M_{\star}\left|\frac{V\Delta}{Pf}\right|\pi j^{2/3}(j-1)^{1/3},
\label{mn}
\end{eqnarray}
As shown in figure 10 of LXW12,  planets' true masses are likely less than the nominal masses. However, one should note that the nominal mass is not a upper limit. In order to further derive the upper mass limit statistically, Paper I used a Monte Carlo method to map the mass and eccentricity by considering the constraints from the TTV amplitude and phase for each planet candidate.

Here we improve on the above Monte Carlo method by replacing it with a direct MCMC fitting to the TTV data using equations \ref{ttveqn} and \ref{vv}.  In such a fitting process, $T$, $T^{'}$, $P$, $P^{'}$, $\mu$, $\mu^{'}$,  ${\rm Re}Z_{\rm free}$, ${\rm Im}Z_{\rm free}$ (Real and imaginal parts of $Z_{\rm free}$ respectively) are 8 fitting parameters. The prior distributions of $\mu$ and $\mu^{'}$ are uniform between 0 and 0.1, while the prior distributions of ${\rm Re}Z_{\rm free}$ and ${\rm Im}Z_{\rm free}$ satisfy the condition that $|Z_{\rm free}|$ is uniform between 0 and 0.8 and its phase is uniform between 0 and 2$\pi$. 
\rr{$|Z_{\rm free}|$ is cut off at 0.8 because extreme large eccentricity ($>$0.2) are very unlikely based on the recent studies on transit durations \citep{Moo11} and TTV phases \citep{WL13}. Larger $|Z_{\rm free}|$ would lead to even smaller mass for the transit objects in favour of confirming planets.}

Figures \ref{fig:ttvfit2a} and \ref{fig:ttvfit2b} plot the MCMC fitting results for all these 15 pairs systems. As expected, the masses are strongly correlated with eccentricity due to the well-known degeneracy between them, and thus generally, one cannot accurately extract individual masses or eccentricities based solely on their TTV \footnote{Sometimes the TTV phases and/or their distribution may help break the degeneracy (see LXW12 and \citet{WL13})}. The posterior mass distribution has a narrow high-end tail, which shows that extremely high masses are possible only in a very narrow parameter space. We then calculate the 99.7\% percentile of the posterior mass distribution and  define it as the 3-$\sigma$ mass upper limit ($m_{\rm max}$, listed in table \ref{tab:planet}). As can be seen, all the 30 candidates have a maximum mass less than 25 Jupiter masses ($M_{\rm J}$) or $< 7945$ Earth masses ($M_{\rm E}$), confirming their planetary nature (Schneider et al. 2011). Furthermore,  as can be seen from the right columns of figures \ref{fig:ttvfit2a} and \ref{fig:ttvfit2b}, TTVs also place a \emph{relatively} tight constraint on the mass ratios of these planet pairs. 

\rr{We also performed another similar set of MCMC fitting but with a logarithmic priors for both planet masses, $\mu$, $\mu^{'}$ and eccentricity $Z_{\rm free}$. We found the results are similar to those shown in figures \ref{fig:ttvfit2a} and \ref{fig:ttvfit2b}.}

\subsection{Other diagnoses}
Although a pair of sinusoidal TTVs provide strong evidence that the TTV signal arises from a pair of interacting planets, rather than from some other astrophysical false positive\footnote{Such as two planets orbiting two different background stars or planets orbiting in a binary system, as discussed in \citet{Ste13}}, we perform two further checks to verify the ``confirmed'' status of these planetary systems.

First, we check the centroid offsets of these targets during transit \citep{Bry13}. If a pair of transits were due to two planets each orbiting different stars, then each planet transit would cause some offset to the target centroid, and the offsets caused by different planets would be in different directions. In all cases studied here, we do not find any significant such offset (by checking the Kepler pipeline data validation report$^{1}$ of each candidate and personal communication with Ji Wang), which is consistent with each pair of planets orbiting the same star. 

Second, we calculate the normalized transit duration ratio, $\xi=(T_{\rm dur}/T_{\rm dur}^{'})(P^{'}/P)^{1/3}$ \citep{Fab12b}, for each pair. The value of $\xi$ should be order of unity if the pair of planets orbits the same star. Indeed, in all cases as listed in table \ref{tab:planet},  it is consistent with each pair of planets being in the same system.

\section{Summary}
\begin{figure}
\begin{center}
\includegraphics[width=\textwidth]{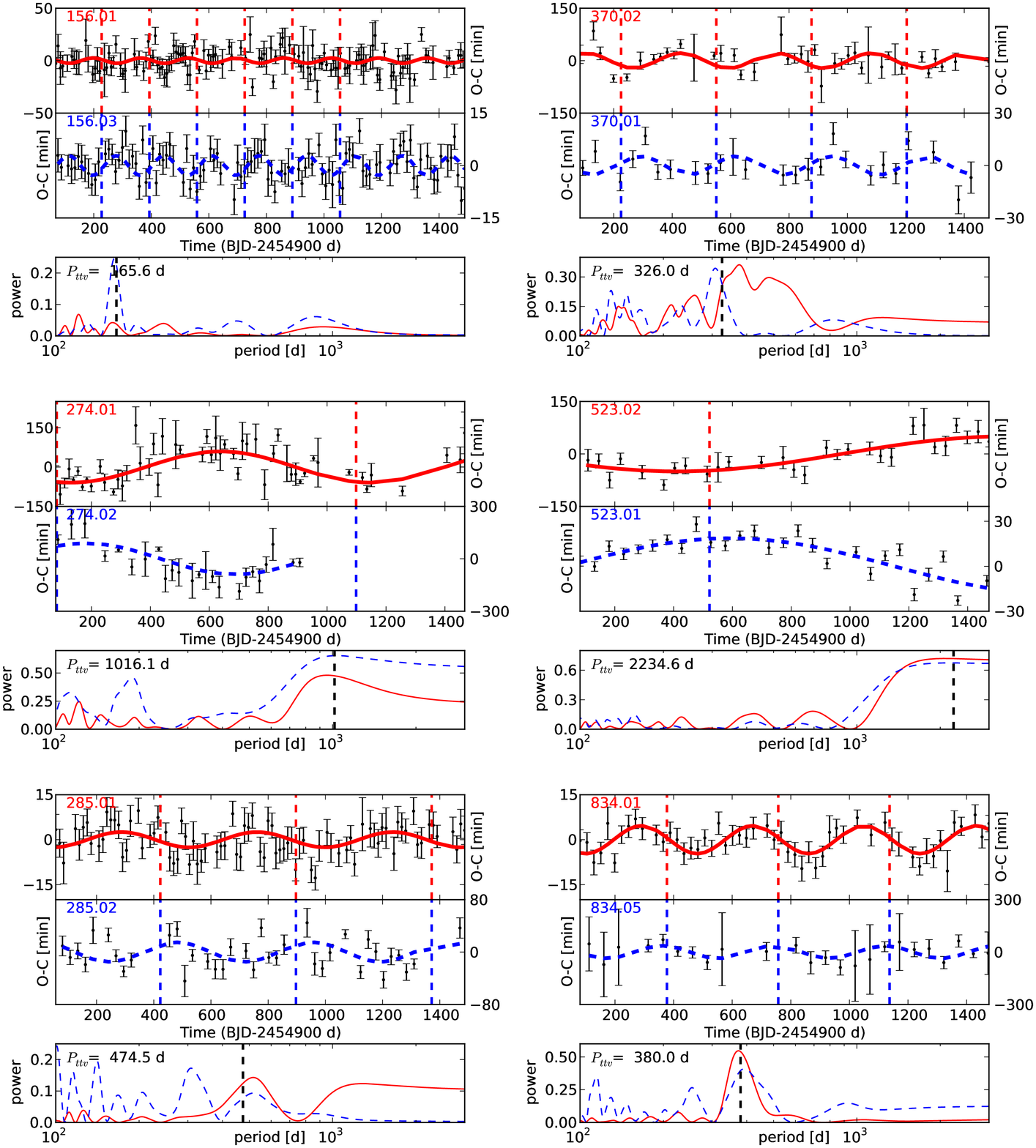}
\caption{TTV evidence for six planet pairs (red lines for the inner one, and blue for the outer one). For each of them, we plot the best-fit theoretical  curves on top of the TTV data and the TTV periodogram.  In the TTV fitting panels, the vertical dashed lines  denote the times when the longitude of conjunction points at the observer, i.e., $\lambda^{j}=0$. In the periodogram panel, the vertical dashed line denotes the theoretically predicted period of the TTV. The TTV fitting results are summarized in table \ref{tab:ttvfit}.}
\label{fig:ttv1}
   \end{center}
\end{figure}

\begin{figure}
\begin{center}
\includegraphics[width=\textwidth]{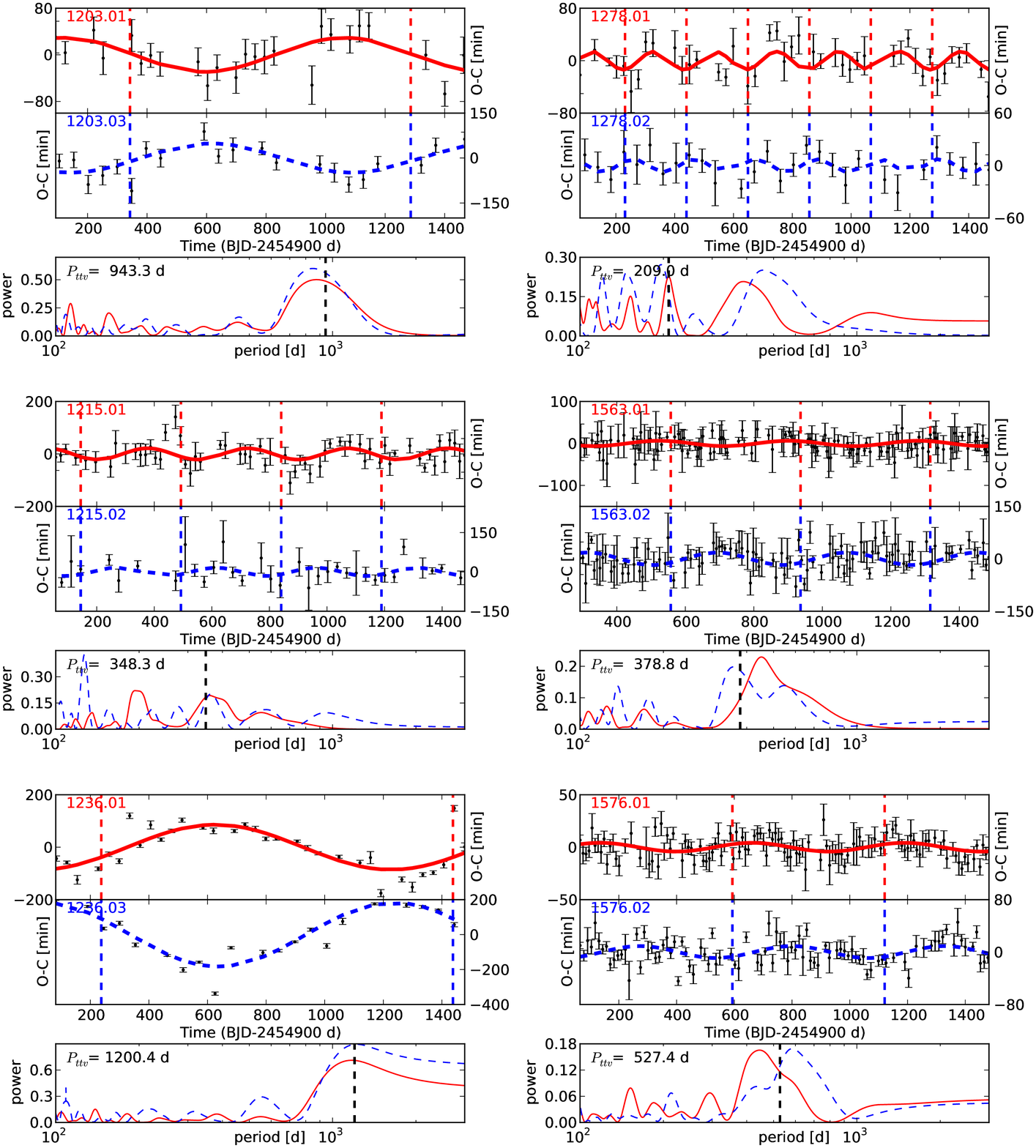}
\caption{\rev{Similar to Fig.\ref{fig:ttv1}, but for another six planet pairs. }}
\label{fig:ttv2}
   \end{center}
\end{figure}

\begin{figure}
\begin{center}
\includegraphics[width=\textwidth]{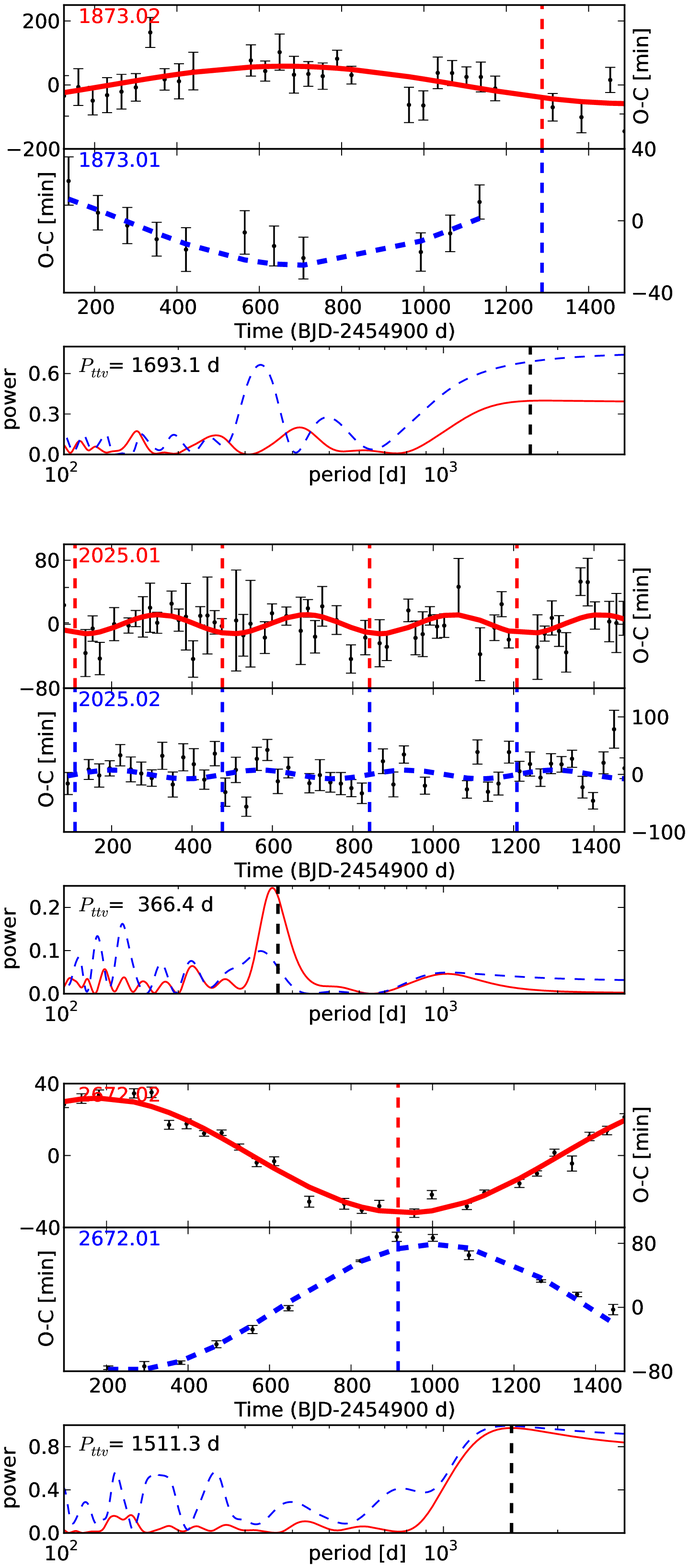}
\caption{\rev{Similar to Fig.\ref{fig:ttv1}, but for another three planet pairs. }}
\label{fig:ttv3}
   \end{center}
\end{figure}

\begin{figure}
\begin{center}
\includegraphics[width=\textwidth]{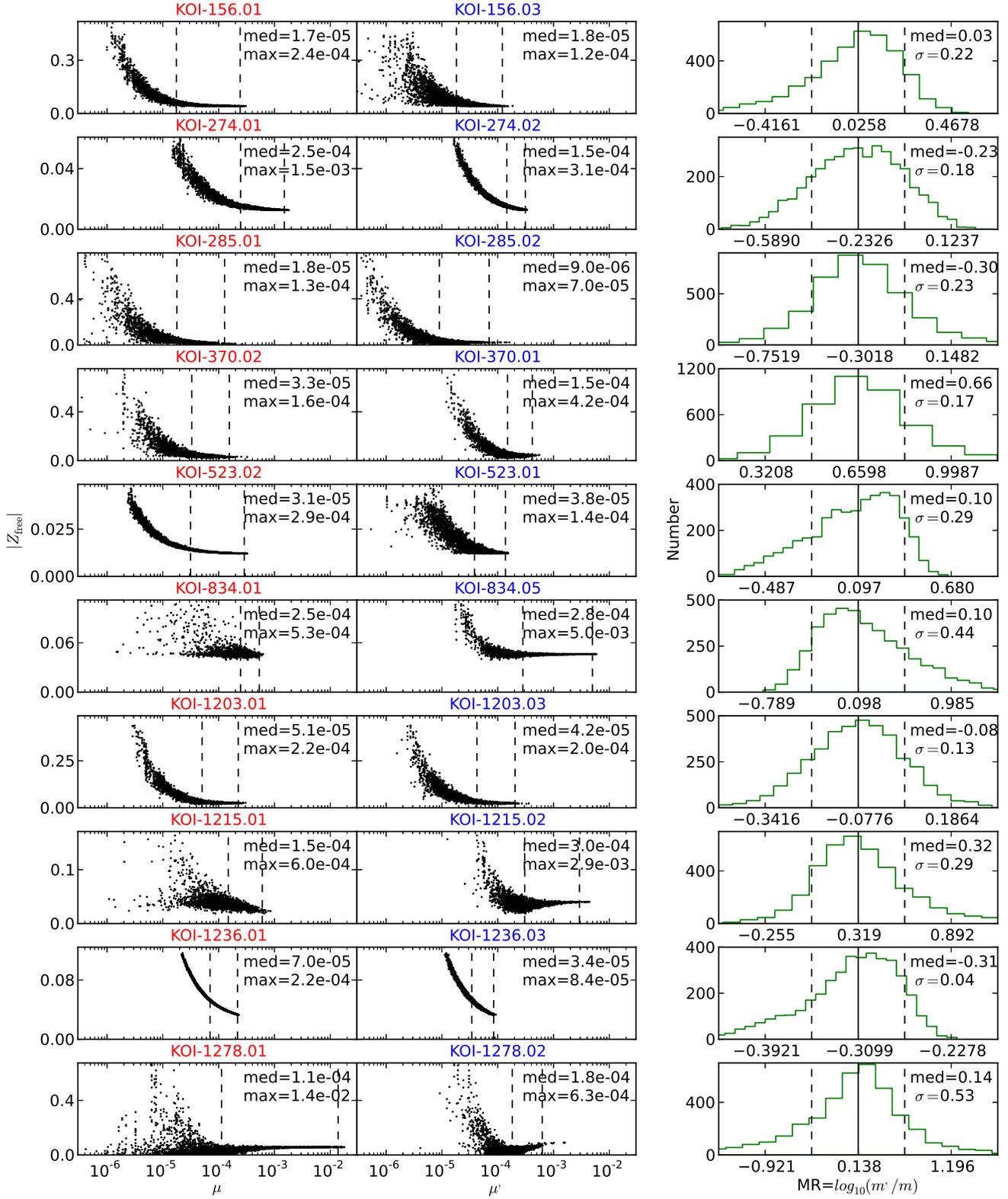}
\caption{MCMC fitting results (section \ref{sec:mass}) to constrain the masses of 10 KOI pairs. The left two columns show the posterior distributions of $\mu$ ($\mu^{'}$) vs. $|Z_{\rm free}|$ with the dashed lines marking the 50\% and 99.7\% percentiles (corresponding to $m_{\rm max}$).  The right column panels show the histogram of the mass ratio index, $MR=log_{10}(m^{'}/m)$, with the vertical lines marking the median and 1-$\sigma$ positions. As expected, mass is highly degenerate with eccentricity. Nevertheless, the mass ratios of the KOI pairs can be \emph{relatively} well constrained.}
\label{fig:ttvfit2a}
   \end{center}
\end{figure}

\begin{figure}
\begin{center}
\includegraphics[width=\textwidth]{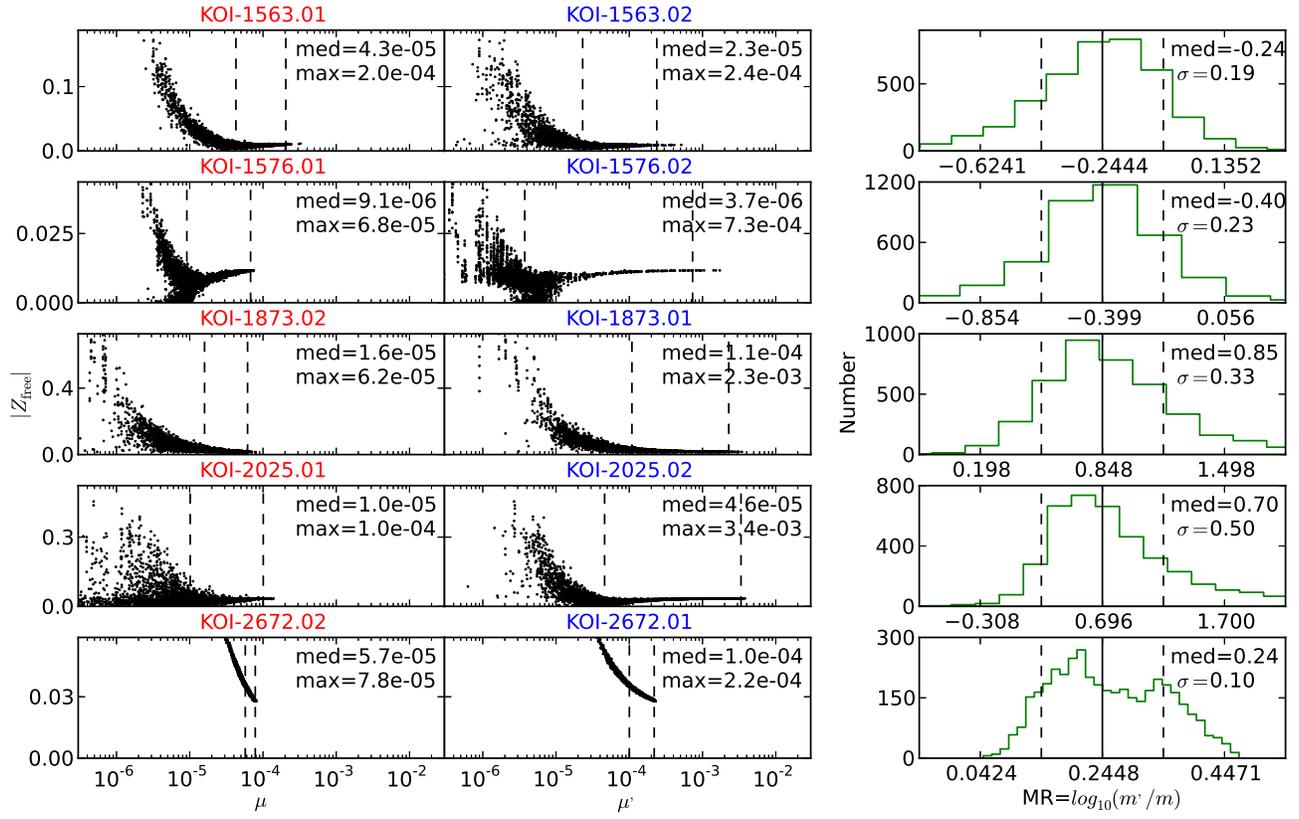}
\caption{Same as figure \ref{fig:ttvfit2a}, but for another 5 KOI pairs.}
\label{fig:ttvfit2b}
   \end{center}
\end{figure}

Expanding on the work of Paper I, we have measured and analyzed the TTVs of a further 15 Kepler planetary candidate pairs. We demonstrate that all 15 systems have measured pairs of TTV signals that agree with the theoretically predicted sinusoidal curves to a high degree of confidence (Fig.\ref{fig:ttv1}, \ref{fig:ttv2} and \ref{fig:ttv3}, and Table \ref{tab:ttvfit}), and that the TTV phases and amplitudes constrain the masses of the objects to lie within the planetary range, leading to the confirmation of 30 planets in 15 planetary systems (Table \ref{tab:planet}).

\begin{figure}
\begin{center}
\includegraphics[width=\textwidth]{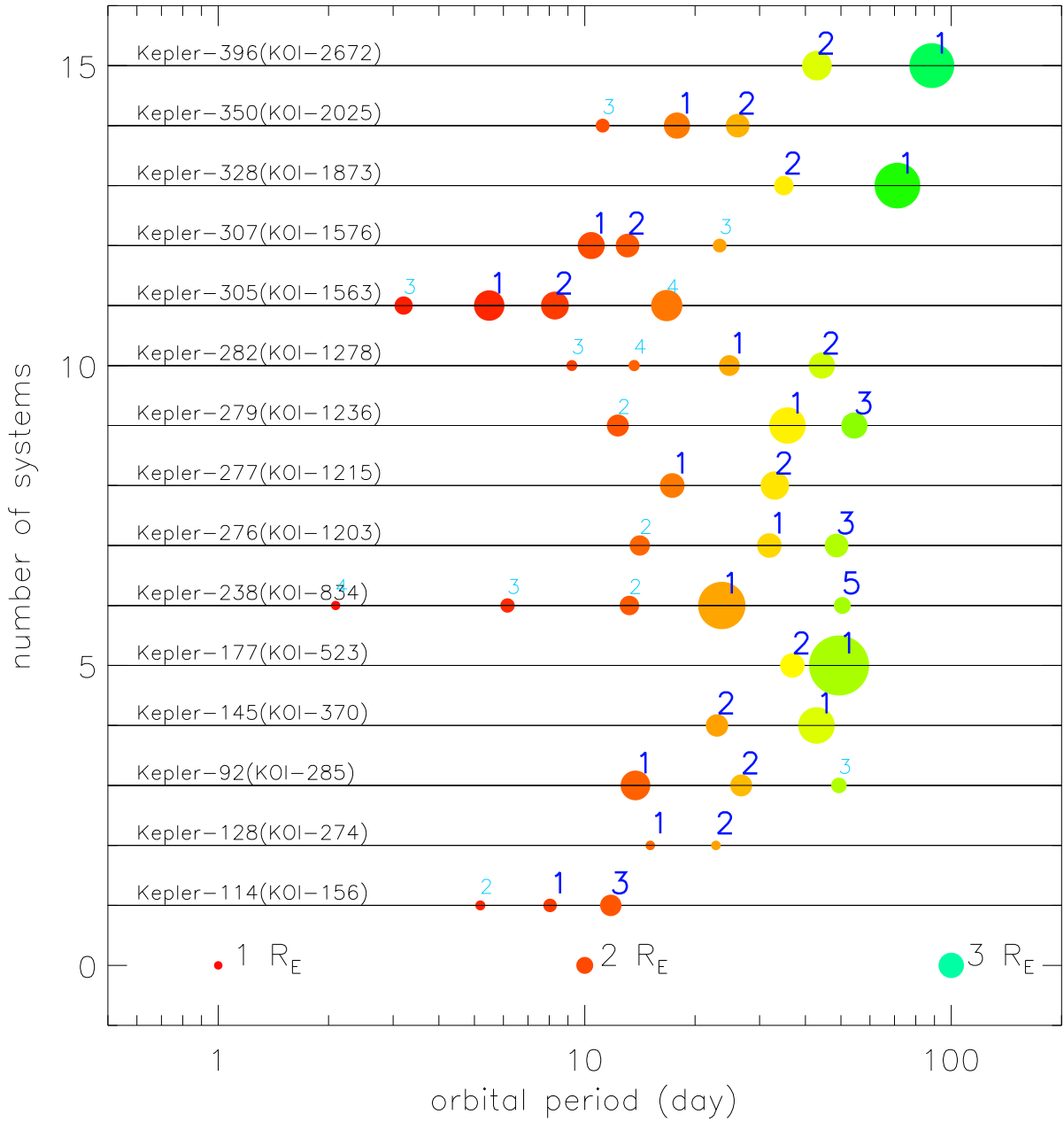}
\caption{15 multiple transiting systems studied in this paper. Planets and candidates are plotted and coloured in order of orbital period. The number beside each planet (in larger blue font, confirmed in this paper) and candidate (in smaller cyan font) is the KOI sequence id. }
\label{fig_sys}
   \end{center}
\end{figure}

In figure \ref{fig_sys}, we plot these 15 systems ordered according to their orbital periods and radii. Many of these systems are multiple systems with 3-5 transiting planet candidates. The two planets confirmed using TTVs in each system are usually the ones with larger sizes and orbital periods, which is consistent with our expectations as TTV amplitudes and their measurement accuracy generally increase with planet size (mass) and orbital period.  Some of these systems are relatively bright (KOI-274, KOI-285, KOI-370 and KOI-2672) with Kepler magnitude less than 12, which are suitable for further follow-up observations.   

\acknowledgments 
JWX thanks 
the referee for a constructive review report, 
Yanqin Wu and Yoram Lithwick for helpful conversations on TTV data analysis,  Ji Wang for assistance in double checking the centroid offset of the transit targets, John Dubinski for providing access to the Sunnyvale cluster at CITA on which most calculation of this work was performed and Matthew Payne for reading and revising the manuscript. JWX acknowledges support from the Ontario government, University of Toronto, the Key Development Program of Basic Research of China (973 program, No. 2013CB834900),   National Natural Science Foundation of China (No. 10925313), 985 Project of Ministration of Education, Superiority Discipline Construction Project of Jiangsu Province. This work would not be done without the beautiful light curves produced by the Kepler team.



\clearpage
\begin{table*}[]
 \begin{center}
  \caption{Transit time measurements of the 30 (15 pairs) newly confirmed planets. }
  \label{tab:ttvdata}
\input{tb_ttv.tex}

\end{center}
\end{table*}

\begin{table*}[]
  \caption{Results of TTV fitting for the 15 Pairs of Planets}
  \label{tab:ttvfit}
\input{tb_all_good_1.tex}
\end{table*}

\begin{table*}[]
  \caption{Key Properties of Planets and Stars of the 15 Systems.}
  \label{tab:planet}
\input{tb_all_good_2.tex}
\end{table*}

\clearpage
\clearpage

\clearpage

\clearpage

\end{CJK*}

\end{document}

%% file: tb_ttv.tex
\begin{tabular}{cccc}
\hline
\hline
$^{a}$ KOI 156.01  &      Ntr=      &   128 \\ 
           n          &       t (d)                 &   et (d) \\
          -1          &     67.9956741      &    0.0034053\\
           0          &     76.0486450      &    0.0076134\\         
           1          &     84.0753174      &    0.0108287\\
          . . .        &     . . .                    &    . . .   \\
         175          &   1483.2742920   &   0.0054487\\
 KOI 156.03  &      Ntr=      &   103 \\ 
           n          &       t (d)                 &   et (d) \\
           0         &    75.7076797         &   0.0033197\\
           1         &    87.4814377         &  0.0025161\\
           3         &    111.0340347      &   0.0020649\\
          . . .        &     . . .                     &    . . .   \\
        120        & 1488.8378906         &  0.0021947\\
 KOI 274.01  &      Ntr=      &   52 \\      
           n          &       t (d)                 &   et (d) \\      
          . . .       &     . . .                  &    . . .   \\     
\hline
\hline\\
\end{tabular}\\
$^{a}$ Here $n$ is the transit sequence id, $\rm t= BJD - 2454900$ d is the transit time with its uncertainty, ${\rm et}$, and Ntr is the number of transits for each candidate. (This table will be available in its entirety in a machine-readable form. A portion is shown here for guidance regarding its form and content.)

%% file: tb_all_good_1.tex
\begin{center}
\renewcommand{\arraystretch}{1.8}
\begin{tabular}{|lr|rrr|rrrr|r|}
\hline
Kepler & KOI & $j$ & $\Delta$ & $P^{j}$ & $A_{ttv}$ & $A^{'}_{ttv}$ & $\phi_{ttv}$ & $\phi^{'}_{ttv}$ & $\rm FAP$ \\
- &  -   &   - & -      & d    & d       & d             & deg          & deg            & -            \\
\hline
114 c-d&156{.01}-{03}&       3& -0.024&   165.6&$    0.0018_{-    0.0008}^{+    0.0008}$&$    0.0019_{-    0.0003}^{+    0.0003}$&$   -25.0_{-    25.8}^{+    25.8}$&$   159.0_{-    10.4}^{+    10.4}$&$< 10^{-4}$\\
128 b-c&274{.01}-{02}&       3&  0.007&  1016.4&$    0.0421_{-    0.0033}^{+    0.0033}$&$    0.0626_{-    0.0116}^{+    0.0116}$&$   -73.4_{-     5.3}^{+     5.3}$&$   124.4_{-     8.1}^{+     8.1}$&    0.0001\\
92 b-c&285{.01}-{02}&       2& -0.028&   474.5&$    0.0018_{-    0.0005}^{+    0.0005}$&$    0.0103_{-    0.0021}^{+    0.0021}$&$    13.4_{-    16.5}^{+    16.5}$&$   226.1_{-    10.3}^{+    10.3}$&    0.0036\\
145 b-c&370{.02}-{01}&       2& -0.066&   326.0&$    0.0150_{-    0.0029}^{+    0.0029}$&$    0.0037_{-    0.0011}^{+    0.0011}$&$    42.7_{-     9.1}^{+     9.1}$&$   194.3_{-    16.9}^{+    16.9}$&    0.0027\\
177 b-c&523{.02}-{01}&       4&  0.006&  2234.6&$    0.0348_{-    0.0141}^{+    0.0141}$&$    0.0130_{-    0.0015}^{+    0.0015}$&$  -107.6_{-    14.9}^{+    14.9}$&$   101.2_{-    11.0}^{+    11.0}$&$< 10^{-4}$\\
238 e-f&834{.01}-{05}&       2&  0.066&   380.0&$    0.0033_{-    0.0005}^{+    0.0005}$&$    0.0249_{-    0.0097}^{+    0.0097}$&$     8.2_{-     8.2}^{+     8.2}$&$    75.8_{-    24.4}^{+    24.4}$&$< 10^{-4}$\\
276 c-d&1203{.01}-{03}&       3&  0.017&   943.3&$    0.0203_{-    0.0047}^{+    0.0047}$&$    0.0338_{-    0.0058}^{+    0.0058}$&$     5.6_{-    14.4}^{+    14.4}$&$   194.1_{-    10.4}^{+    10.4}$&$< 10^{-4}$\\
277 b-c&1215{.01}-{02}&       2& -0.047&   348.3&$    0.0150_{-    0.0034}^{+    0.0034}$&$    0.0111_{-    0.0050}^{+    0.0050}$&$    29.5_{-    14.9}^{+    14.9}$&$   143.4_{-    22.3}^{+    22.3}$&    0.0008\\
279 c-d&1236{.01}-{03}&       3&  0.015&  1200.4&$    0.0596_{-    0.0015}^{+    0.0015}$&$    0.1276_{-    0.0018}^{+    0.0018}$&$  -153.4_{-     1.4}^{+     1.4}$&$    29.6_{-     0.8}^{+     0.8}$&$< 10^{-4}$\\
282 d-e&1278{.01}-{02}&       2& -0.106&   209.0&$    0.0097_{-    0.0026}^{+    0.0026}$&$    0.0054_{-    0.0031}^{+    0.0031}$&$   100.8_{-    13.4}^{+    13.4}$&$   223.7_{-    29.0}^{+    29.0}$&    0.0013\\
305 b-c&1563{.01}-{02}&       3&  0.007&   378.8&$    0.0044_{-    0.0016}^{+    0.0016}$&$    0.0124_{-    0.0022}^{+    0.0022}$&$    57.8_{-    23.8}^{+    23.8}$&$   219.0_{-    11.9}^{+    11.9}$&    0.0004\\
307 b-c&1576{.01}-{02}&       5&  0.005&   527.5&$    0.0025_{-    0.0006}^{+    0.0006}$&$    0.0065_{-    0.0008}^{+    0.0008}$&$    99.7_{-    15.4}^{+    15.4}$&$   226.1_{-     7.2}^{+     7.2}$&$< 10^{-4}$\\
328 b-c&1873{.02}-{01}&       2&  0.021&  1693.1&$    0.0407_{-    0.0117}^{+    0.0117}$&$    0.0172_{-    0.0070}^{+    0.0070}$&$   -42.8_{-    23.8}^{+    23.8}$&$   144.6_{-    45.0}^{+    45.0}$&$< 10^{-4}$\\
350 c-d&2025{.01}-{02}&       3& -0.024&   366.4&$    0.0083_{-    0.0024}^{+    0.0024}$&$    0.0053_{-    0.0027}^{+    0.0027}$&$    65.3_{-    15.3}^{+    15.3}$&$   178.9_{-    30.7}^{+    30.7}$&    0.0006\\
396 b-c&2672{.02}-{01}&       2&  0.029&  1511.3&$    0.0222_{-    0.0006}^{+    0.0006}$&$    0.0549_{-    0.0017}^{+    0.0017}$&$   -85.4_{-     1.5}^{+     1.5}$&$   111.4_{-     1.5}^{+     1.5}$&$< 10^{-4}$\\
\hline
\end{tabular} \\
\end{center}
Note: systems KOI-156 and KOI-1215 have also been analyzied by \citet{WL13} using the TTV data from Q0 to Q6. \\

%% file: tb_all_good_2.tex
\begin{center}
\renewcommand{\arraystretch}{2.2}
\resizebox{\columnwidth}{!}{%
\begin{tabular}{|lr|rrrrrrrr|rrrrr|}
\hline
Kepler & KOI & $P$$^{a}$ & $P^{'}$$^{a}$ & $R$$^{b}$  & $R^{'}$$^{b}$ & $m_{\rm nom}$$^{c}$ & $m^{'}_{\rm nom}$$^{c}$ & $m_{\rm max}$$^{d}$ & $m^{'}_{\rm max}$$^{d}$ & mag$^{e}$ & $M_{\star}$$^{f}$ & log(g)$^{g}$ & $R_{\star}$$^{g}$ &  $\xi$$^{h}$ \\
 -  & d   &  d      &$R_{\rm E}$ & $R_{\rm E}$   & $M_{\rm E}$         & $M_{\rm E}$             & $M_{\rm E}$         & $M_{\rm E}$             & -& $M_{\odot}$    &-  & $R_{\odot}$       &  -           \\
\hline
114 c-d&156{.01}-{03}&   8.041&  11.776&$1.60_{-0.18}^{+0.18}$&$2.53_{-0.28}^{+0.28}$&$     2.8_{-     0.6}^{+     0.6}$&$     3.9_{-     1.7}^{+     1.7}$&    45.3&    22.6&13.7&$  0.56_{-  0.06}^{+  0.06}$&$4.72_{-0.07}^{+0.07}$&$0.54_{-0.06}^{+0.06}$&1.03\\
128 b-c&274{.01}-{02}&  15.090&  22.804&$1.13_{-0.03}^{+0.03}$&$1.13_{-0.03}^{+0.03}$&$    30.7_{-     6.0}^{+     6.0}$&$    33.3_{-     3.3}^{+     3.3}$&   589.8&   122.9&11.4&$  1.18_{-  0.07}^{+  0.07}$&$4.07_{-0.01}^{+0.01}$&$1.66_{-0.04}^{+0.04}$&0.93\\
92 b-c&285{.01}-{02}&  13.749&  26.723&$3.51_{-0.10}^{+0.10}$&$2.60_{-0.08}^{+0.08}$&$    64.3_{-    13.9}^{+    13.9}$&$     6.1_{-     1.8}^{+     1.8}$&    51.4&    28.3&11.6&$  1.21_{-  0.08}^{+  0.08}$&$4.05_{-0.01}^{+0.01}$&$1.70_{-0.05}^{+0.05}$&0.89\\
145 b-c&370{.02}-{01}&  22.951&  42.882&$2.65_{-0.08}^{+0.08}$&$4.32_{-0.12}^{+0.12}$&$    37.1_{-    11.6}^{+    11.6}$&$    79.4_{-    16.4}^{+    16.4}$&    68.3&   183.1&11.9&$  1.32_{-  0.10}^{+  0.10}$&$4.02_{-0.01}^{+0.01}$&$1.85_{-0.05}^{+0.05}$&0.59\\
177 b-c&523{.02}-{01}&  36.855&  49.412&$2.90_{-0.30}^{+1.52}$&$7.10_{-0.72}^{+3.71}$&$     2.0_{-     0.3}^{+     0.5}$&$     7.5_{-     3.1}^{+     3.5}$&   102.4&    49.1&15.0&$  1.07_{-  0.12}^{+  0.25}$&$4.41_{-0.32}^{+0.06}$&$1.07_{-0.11}^{+0.56}$&1.58\\
238 e-f&834{.01}-{05}&  23.654&  50.447&$5.60_{-0.46}^{+2.37}$&$2.00_{-0.17}^{+0.85}$&$   169.7_{-    69.1}^{+    71.7}$&$    13.5_{-     2.5}^{+     2.9}$&   188.5&  1758.5&15.1&$  1.06_{-  0.12}^{+  0.17}$&$4.49_{-0.30}^{+0.04}$&$0.96_{-0.08}^{+0.41}$&1.46\\
276 c-d&1203{.01}-{03}&  31.884&  48.648&$2.90_{-0.28}^{+1.27}$&$2.80_{-0.27}^{+1.23}$&$    16.6_{-     3.6}^{+     4.4}$&$    16.3_{-     4.3}^{+     5.0}$&    82.2&    74.9&15.4&$  1.10_{-  0.14}^{+  0.22}$&$4.46_{-0.30}^{+0.05}$&$1.03_{-0.10}^{+0.45}$&1.62\\
277 b-c&1215{.01}-{02}&  17.324&  33.006&$2.92_{-0.63}^{+0.73}$&$3.36_{-0.72}^{+0.83}$&$    87.3_{-    39.9}^{+    41.7}$&$    64.2_{-    15.7}^{+    18.1}$&   225.2&  1085.7&13.4&$  1.12_{-  0.11}^{+  0.19}$&$4.03_{-0.13}^{+0.17}$&$1.69_{-0.36}^{+0.42}$&1.27\\
279 c-d&1236{.01}-{03}&  35.736&  54.414&$4.30_{-0.41}^{+1.72}$&$3.10_{-0.30}^{+1.24}$&$    49.4_{-     5.9}^{+     7.2}$&$    37.5_{-     4.5}^{+     5.5}$&    79.7&    31.0&13.7&$  1.10_{-  0.13}^{+  0.16}$&$4.42_{-0.28}^{+0.05}$&$1.07_{-0.10}^{+0.43}$&1.20\\
282 d-e&1278{.01}-{02}&  24.806&  44.347&$2.46_{-0.20}^{+1.00}$&$3.10_{-0.25}^{+1.26}$&$    61.0_{-    36.1}^{+    35.9}$&$    56.2_{-    16.7}^{+    16.2}$&  4462.8&   202.6&15.2&$  0.97_{-  0.12}^{+  0.10}$&$4.53_{-0.30}^{+0.03}$&$0.89_{-0.07}^{+0.36}$&1.01\\
305 b-c&1563{.01}-{02}&   5.487&   8.291&$3.60_{-0.36}^{+0.90}$&$3.30_{-0.33}^{+0.82}$&$    10.5_{-     2.0}^{+     2.6}$&$     6.0_{-     2.2}^{+     2.4}$&    51.9&    60.1&15.8&$  0.76_{-  0.06}^{+  0.13}$&$4.51_{-0.26}^{+0.07}$&$0.80_{-0.08}^{+0.20}$&1.06\\
307 b-c&1576{.01}-{02}&  10.416&  13.084&$3.20_{-0.46}^{+1.20}$&$2.80_{-0.41}^{+1.05}$&$     3.1_{-     0.5}^{+     0.6}$&$     1.5_{-     0.4}^{+     0.5}$&    22.1&   239.2&14.1&$  0.98_{-  0.09}^{+  0.14}$&$4.36_{-0.25}^{+0.12}$&$1.08_{-0.16}^{+0.41}$&0.97\\
328 b-c&1873{.02}-{01}&  34.921&  71.312&$2.30_{-0.23}^{+0.96}$&$5.40_{-0.54}^{+2.24}$&$    28.5_{-    12.3}^{+    12.9}$&$    39.4_{-    12.6}^{+    13.6}$&    23.6&   874.5&15.7&$  1.15_{-  0.16}^{+  0.22}$&$4.44_{-0.27}^{+0.05}$&$1.06_{-0.11}^{+0.44}$&0.84\\
350 c-d&2025{.01}-{02}&  17.849&  26.136&$3.10_{-0.60}^{+1.42}$&$2.80_{-0.54}^{+1.28}$&$     6.1_{-     3.2}^{+     3.3}$&$    14.9_{-     4.7}^{+     5.3}$&    33.6&  1117.4&13.8&$  1.00_{-  0.12}^{+  0.20}$&$4.24_{-0.25}^{+0.18}$&$1.25_{-0.24}^{+0.57}$&0.98\\
396 b-c&2672{.02}-{01}&  42.994&  88.505&$3.50_{-0.65}^{+1.28}$&$5.30_{-0.99}^{+1.95}$&$    75.5_{-     5.8}^{+    11.8}$&$    17.9_{-     1.3}^{+     2.8}$&    22.2&    62.0&11.9&$  0.85_{-  0.06}^{+  0.13}$&$4.32_{-0.23}^{+0.18}$&$1.06_{-0.20}^{+0.39}$&0.87\\
\hline
\end{tabular} } \\
\end{center}
$^{a}$ Their uncertainties are all less than $10^{-3}$ d\\
$^{b}$ Their errorbars reflect the uncertainties of their transit lightcurve fittings (i.e., $R_{\rm p}/R_{\star}$) and stellar radii.\\
$^{c}$ Derived from Eqn.\ref{mn}. Their errorbars reflect the uncertainties of their TTV emplitudes and stellar masses.\\
$^{d}$ Defined as the 99.7\% percentile of the posterior mass distribution from the MCMC fitting (section 2.4) \\
$^{e}$ Kepler magnitude.\\
$^{f,g}$ Stellar properties are adopted from the revised Q1-Q16 catalog \citep{Hub13},which are available at the NASA exoplanet archive (http://exoplanetarchive.ipac.caltech.edu). \\
$^{h}$ The normalized transits duration ratio, $\xi=(T_{\rm dur}/T_{\rm dur}^{'})(P^{'}/P)^{1/3}$ \citep{Fab12b}. \\